# Chemical heterogeneity at conducting ferroelectric domain walls

**Kasper A. Hunnestad[1,2,*], Guo-Dong Zhao[3], Mao-Hua Zhang[3], Tiannan Yang[3], Elzbieta Gradauskaite[4,5], Antonius T. J. van Helvoort[6], Morgan Trassin[4], Long-Qing Chen[3], Tadej Rojac[7], Dennis Meier[1,8,9,*]**

*[1]Department of Materials Science and Engineering, Norwegian University of Science and Technology (NTNU), 7491 Trondheim, Norway*
*[2]Present address: Department of Electronic Systems, Norwegian University of Science and Technology (NTNU), Trondheim, Norway*
*[3]Department of Materials Science and Engineering, The Pennsylvania State University, University Park, PA, 16802, USA*
*[4]Department of Materials, ETH Zurich, 8093 Zurich, Switzerland*
*[5]Unité Mixte de Physique, CNRS, Thales, Université Paris-Sud, Université Paris-Saclay, Palaiseau, France*
*[6]Department of Physics, Norwegian University of Science and Technology (NTNU), 7491 Trondheim, Norway*
*[7]Electronic Ceramics Department, Jožef Stefan Institute, 1000 Ljubljana, Slovenia.*
*[8]Faculty of Physics and Center for Nanointegration Duisburg-Essen (CENIDE), University of Duisburg-Essen, Duisburg, Germany*
*[9]Research Center Future Energy Materials and Systems, Research Alliance Ruhr, 44780 Bochum, Germany*
**hunnestad.kasper@gmail.com; dennis.meier@uni-due.de*

---

**Natural interfaces in ferroic oxides have developed into versatile playgrounds for studying electronic correlation effects in 2D systems. The microscopic origin of the emergent local electronic properties is often debated, however, as quantitative atomic-scale characterization remains challenging. A prime example is enhanced conductivity at ferroelectric domain walls, attributed to mechanisms ranging from local band gap reduction to point defect accumulations. Here, we resolve the microscopic mechanisms for domain wall conduction in the ferroelectric model system $BiFeO_3$, by combining transport measurements with atom probe tomography to quantify the local chemical composition and correlate it with the electrical properties. Significant chemical variations along the walls are observed, demonstrating an outstanding chemical flexibility at domain walls, which manifest in spatially varying physical properties. The results give a unifying explanation for the diverse electronic behavior observed and establish the fundamental notion that multiple conduction mechanisms can coexist within individual domain walls.**

The observation of sheet superconductivity at twin walls in $WO_3$ and enhanced electronic conductance at ferroelectric domain walls in $BiFeO_3$ triggered a constantly growing interest in naturally occurring interfaces in oxides and their functional behaviors[1–5]. It is now established that the emergence of unusual transport properties is a general phenomenon observed at domain walls in various materials[6]. Aside from their intriguing physics, such domain walls hold great promise as ultra-small active elements for nanoelectronics and unconventional computing schemes[7]. Examples range from domain wall-based digital switches and rectifiers to memristive devices for neuromorphic computing. An important breakthrough was the identification of three fundamental mechanisms promoting enhanced electronic conduction at domain walls: polarization charges and bandgap reduction (intrinsic), and ionic point defects (extrinsic)[1,8]. This classification provides a comprehensive understanding of the existing conduction mechanisms but also highlights the complexity of the physics and the challenges for property engineering and optimization towards envisioned applications.

Despite the improved understanding of the general nanoscale physics, the main driving mechanism for domain wall conduction often remains elusive, and even for well-studied model systems it is still debated whether it is determined by intrinsic or extrinsic effects. In $LiNbO_3$ and $Pb(Zr_{0.2}Ti_{0.8})O_3$, for instance, both electronic carriers and oxygen vacancies have been reported as sources for the locally enhanced conductance[9,10]. A substantial number of domain wall studies have been performed on $BiFeO_3$, but also here no consensus is reached[11]. In some cases, local alterations in the electronic bands, polarization discontinuities, or a distortion of the polarization structure at the domain walls were associated with their conduction behavior[3,12,13]. In other cases, it was attributed to point defects occupying intermediate energy states in the band gap, such as bismuth vacancies, oxygen vacancies and electron/holes trapped at Fe sites[14–18]. The different mechanisms can operate individually or in a concerted way, making it even more complicated to contextualize the diverse findings reported.

Here, we quantify the chemical composition at ferroelectric domain walls in the $BiFeO_3$ system, investigating the relation between their electronic conduction and the local concentration of ionic point defects. By performing correlated conducting atomic force microscopy (cAFM) and atom probe tomography (APT), we identify differences in chemical composition between different domain walls within the same material, as well as substantial variations along individual domain walls. These observations are consistent with the spatially varying transport properties, as we elaborate based on phase field simulations. The chemical measurements translate into a zoo of different domain wall structures, ranging from point-defect-free to point-defect-rich walls with various types of defects, including oxygen ($V_O$), bismuth ($V_{Bi}$), and iron ($V_{Fe}$) vacancies. The findings suggest a much higher chemical flexibility of the domain walls than previously assumed, which can be leveraged to tune their functional responses.

## Results and discussion

We begin with a general characterization of ferroelectric domain walls in a $BiFeO_3$ polycrystal (see Methods for information about synthesis). Our focus is on the 109° domain walls at which bismuth vacancies and $Fe^{4+}$ states were reported to accumulate and be associated with a locally enhanced conduction[15,19], making them an ideal model system for studying chemistry-property relations at the domain wall level. Figure 1a shows an image of the ferroelectric domain structure recorded by piezoresponse force microscopy (PFM, out-of-plane contrast). Within the marked grain, stripe-like ferroelectric domains are observed with straight domain walls of several micrometers, consistent with previous reports[20]. Across these domain walls, the polarization changes direction by 109°, as we confirm by vector-PFM (see Supplementary Fig. 1). Corresponding cAFM data is displayed in the inset to Fig. 1a, indicating that every other 109° domain wall exhibits enhanced conductance, analogous to the results reported in Ref. [15]. The atomic-scale structure of the domain walls is presented in Fig. 1b, which shows a representative high-resolution high-angle annular dark-field scanning transmission electron microscopy (HAADF-STEM) image (see methods). Across the domain walls, the relative Fe displacements of the Bi sublattice gradually changes[21], leading to a structural wall width of ∼2 nm.

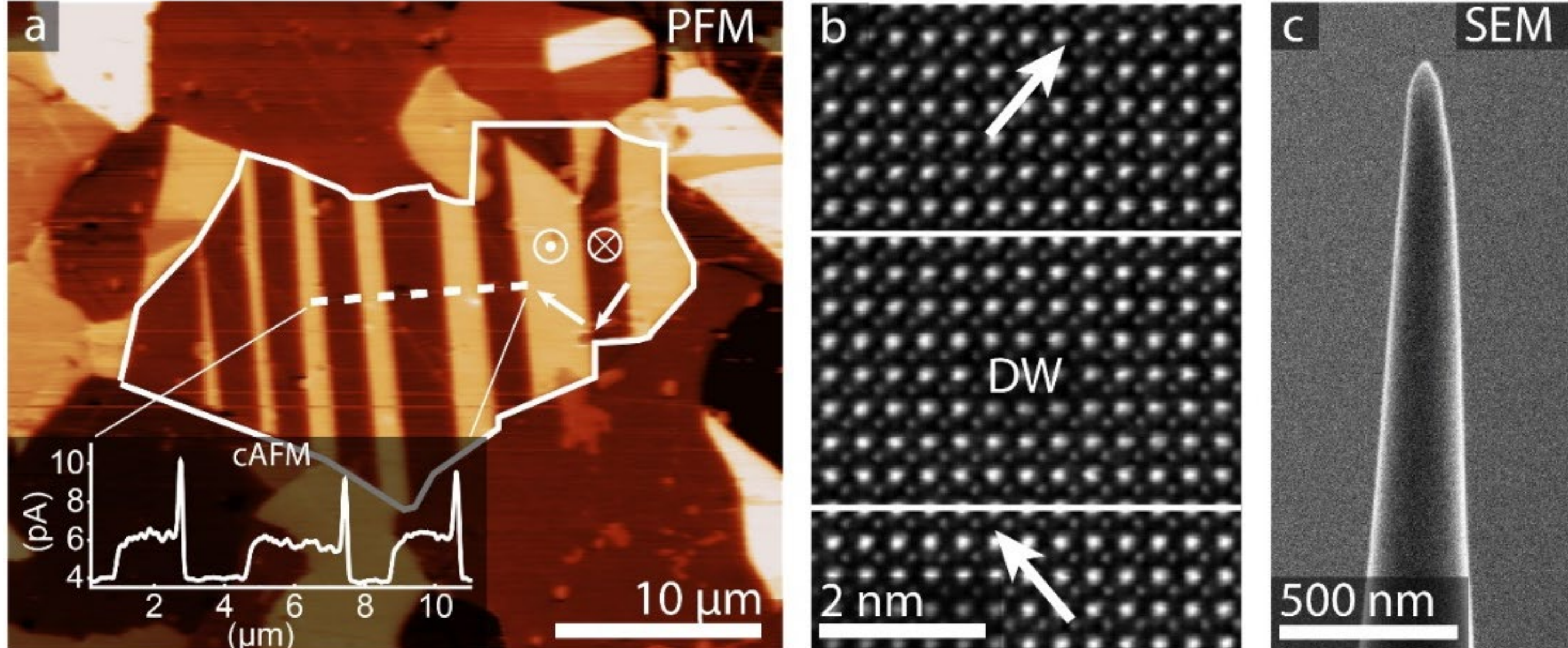

***Fig 1: Domain wall characterization and extraction in $BiFeO_3$. a,*** *Out-of-plane PFM image of a $BiFeO_3$ polycrystal, showing contrast from grains and ferroelectric domains (alternating bright (+P) and dark (-P))[22]. One grain with a characteristic domain structure is outlined in white, and the polarization directions within this grain are labelled. The dashed white line indicates where the cAFM profile is extracted from (inset), which shows enhanced conductivity at the domain walls.* ***b,*** *HAADF-STEM image of a representative 109˚ domain wall in $BiFeO_3$ marked by the solid white lines, giving an estimate for the domain wall width. The atomic structure and Fe displacements from the Bi sublattice center, which correlates to the polarization orientation, are indicated by the arrows.* ***c,*** *Secondary electron SEM image of an APT needle extracted at a domain wall.*

In order to enable a local chemical analysis of the conducting 109° domain walls by APT, we again map the ferroelectric domains, now using a dual-beam focused ion beam scanning electron microscope (FIB-SEM)[22]. This step allows for the visualization, marking, and site-specific extraction of conducting 109° domain walls for APT analysis (see Methods and Supplementary Fig. 2 for details). Figure 1c presents an example of the specimens extracted for APT analysis, with the typical needle-like shape required for APT analysis[23,24].

To ensure we extract the region of interest and that the prepared sample (Fig. 1c) indeed contains a ferroelectric 109° domain wall, a dark-field TEM (DF-TEM) image is recorded prior to the APT measurement (Fig. 2a). The DF-TEM data shows a dark straight line at which an abrupt change in the contrast fringes can be seen. This line goes diagonally across the image and marks the domain wall position[25,26]. The selected area diffraction (SAD) pattern in the inset is indexed to determine the unit cell orientation and, in combination with PFM, the direction of the polar axes (see also Supplementary Fig. 3). In summary, Fig. 1c and Fig. 2a demonstrate that our approach allows for a targeted extraction of individual domain walls for APT measurements, in this case, the extraction of a ferroelectric 109° domain wall.

Figure 2b presents the three-dimensional (3D) APT reconstruction from the specimen in Fig. 2a, where, for clarity, only the Bi ions are displayed (see Methods for analysis details). From such reconstructions, 3D atomic distributions can be obtained to reveal local fluctuations in chemistry[27,28]. Interestingly, the 3D reconstruction shows an anomaly in the distribution of Bi atoms, seen as a slightly darker region in the 2D projection, as indicated by the white dashed lines. The position and orientation in this viewing direction are identical to the position of the domain wall imaged by DF-TEM, which leads us to the conclusion that the measured local deviation in Bi content is associated with the 109° domain wall.

For quantitative information, we consider the chemical composition profile calculated perpendicular to the domain wall plane (Fig. 2c). Both Bi and O concentration profiles exhibit a local minimum at the position of the domain wall with a decrease of about 0.4 at. %, demonstrating that the domain wall is chemically different from the surrounding domains. In contrast, the Fe concentration shows a pronounced increase. Note, however, that the concentration profiles in Fig. 2c represent relative changes. For example, if one species (A) out of three otherwise equally distributed atomic species (33 % of A, B, and C) was locally missing, the measured concentration of the other two would rise (50 % of B and C). Thus, considering the close-packed perovskite structure of $BiFeO_3$, which favors vacancies over energetically much more costly interstitials, and comparable formation energies for Bi, Fe, and O vacancies[29], we conclude from the measured concentration (Fig. 2c) that oxygen ($V_O$) and bismuth ($V_{Bi}$) vacancies accumulate at the domain wall. Going beyond previous studies[14,15], the APT data gives quantitative insight, showing that the local concentration of $V_O$ and $V_{Bi}$ is increased by 0.4 at.% at the domain wall compared to the neighboring domains.

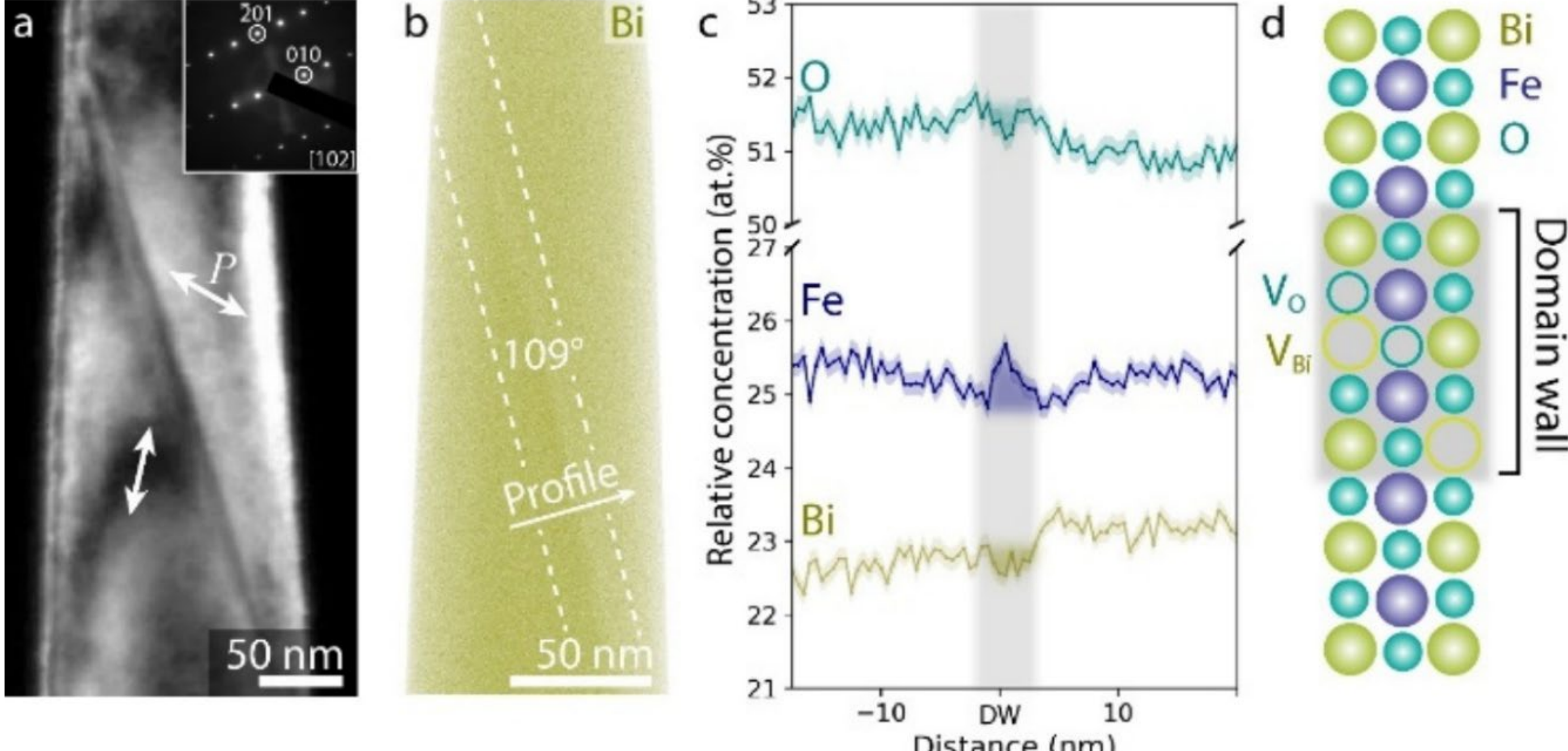

***Fig 2: APT analysis of a ferroelectric domain wall. a,*** *DF-TEM image of a region of an APT specimen containing a single ferroelectric 109 °domain wall, visualized by using the* $\mathbf{0\bar{1}0}$ *reflection. The polarization directions are indicated (see also Supplementary Fig. 3).* ***b,*** *3D APT reconstruction of the 109° domain wall. The domain wall can be seen as a slightly darker line and is marked by the white dashed lines.* ***c,*** *Profile of the chemical composition across the 109 °domain wall (as indicated in* ***b****). An increase in Fe is observed at the domain wall position, indicating an accumulation of* $V_{Bi}$ *and* $V_O$*.* ***d,*** *Schematic of the point defect model based on the APT analysis, containing* $V_{Bi}$ *and* $V_O$*.*

One important question is to what extent the quantitative local results in Fig. 2 are representative for the $BiFeO_3$ system in general. To address this point, we perform additional correlated microscopy measurements at various positions, including the evaluation of the chemical composition along individual 109° domain walls as summarized in Fig. 3. Figure 3a shows another PFM image of the ferroelectric domain structure in the $BiFeO_3$ polycrystal, recorded in a different region than the data in Fig. 1. The PFM map shows an extended 109° domain wall where two positions are marked by white dashed circles. At these positions, which are approximately 6 μm apart, needle-shaped specimens for APT analysis are extracted following the same procedure as for the experiment shown in Fig. 2. Complementary DF-TEM images are shown in Supplementary Fig. 4, corroborating that the targeted 109° domain wall is present in both APT samples. Figure 3b,c present corresponding chemical profiles, mapping the elemental composition across the domain walls in positions ① and ②, respectively.

The position-dependent APT analysis reveals substantial differences in the chemical composition at positions ① and ②. At position ①, we measure a depletion in the O concentration of ≈0.4 at.% and an increase in Bi concentration of ≈0.4 at.%, which is interpreted as an accumulation of both $V_O$ and $V_{Fe}$ vacancies (see Methods), expanding the list of cation vacancies previously observed at $BiFeO_3$ domain walls by including iron[30]. Although Bi anti-site defects ($Bi_{Fe}$) are also a possible interpretation for these chemical composition measurements, additional analyses on $BiFeO_3$ thin films suggest $V_{Fe}$ to be prevalent at $BiFeO_3$ domain walls (see Supplementary Fig. 5)[31]. In contrast, no variation in chemical composition is detected at position ②, within the sensitivity of the experiment (≈0.1 at.%), implying an almost ionic-defect-free domain wall section. Furthermore, it is important to note that the chemical domain wall profiles in Fig. 3a,b are different from the one presented in Fig. 2c. These APT measurements thus establish that ferroelectric 109° domain walls in $BiFeO_3$ do not have a characteristic chemical fingerprint, and even along individual domain walls the chemical composition varies. A more detailed analysis of the compositional variations within a domain wall (from Fig. 2) is presented in Fig. 3d-f. Evaluated in the x- and z-direction within the domain wall plane, as defined in Fig. 3d, we observe that the density of $V_{Bi}$ is heterogeneous, with a total fluctuation of about 0.3 at.% over a distance of about ≈70 nm.

The observed chemical flexibility is remarkable, as it requires a reassessment of the canonical domain wall picture, where only accumulations or depletions of certain types of defects are considered. Our APT data show that the situation is much more complex: Ferroelectric 109° domain walls in $BiFeO_3$ can be free of ionic defects or decorated with all different types of vacancies (i.e., $V_O$, $V_{Fe}$, $V_{Bi}$), with substantial variations in defect density and dominant defect type on the length scale of a few tens of nanometers. Additionally, although the number of samples is still low, we see a tendency that more defect-rich domain walls are formed in the polycrystal than in the thin film investigated here. In total, 11 domain walls from the polycrystal have been investigated out of which four hosted $V_O$, one hosted $V_{Fe}$, four hosted $V_{Bi}$, and seven did not display any measurable level of ionic defects (Supplementary Table 1).

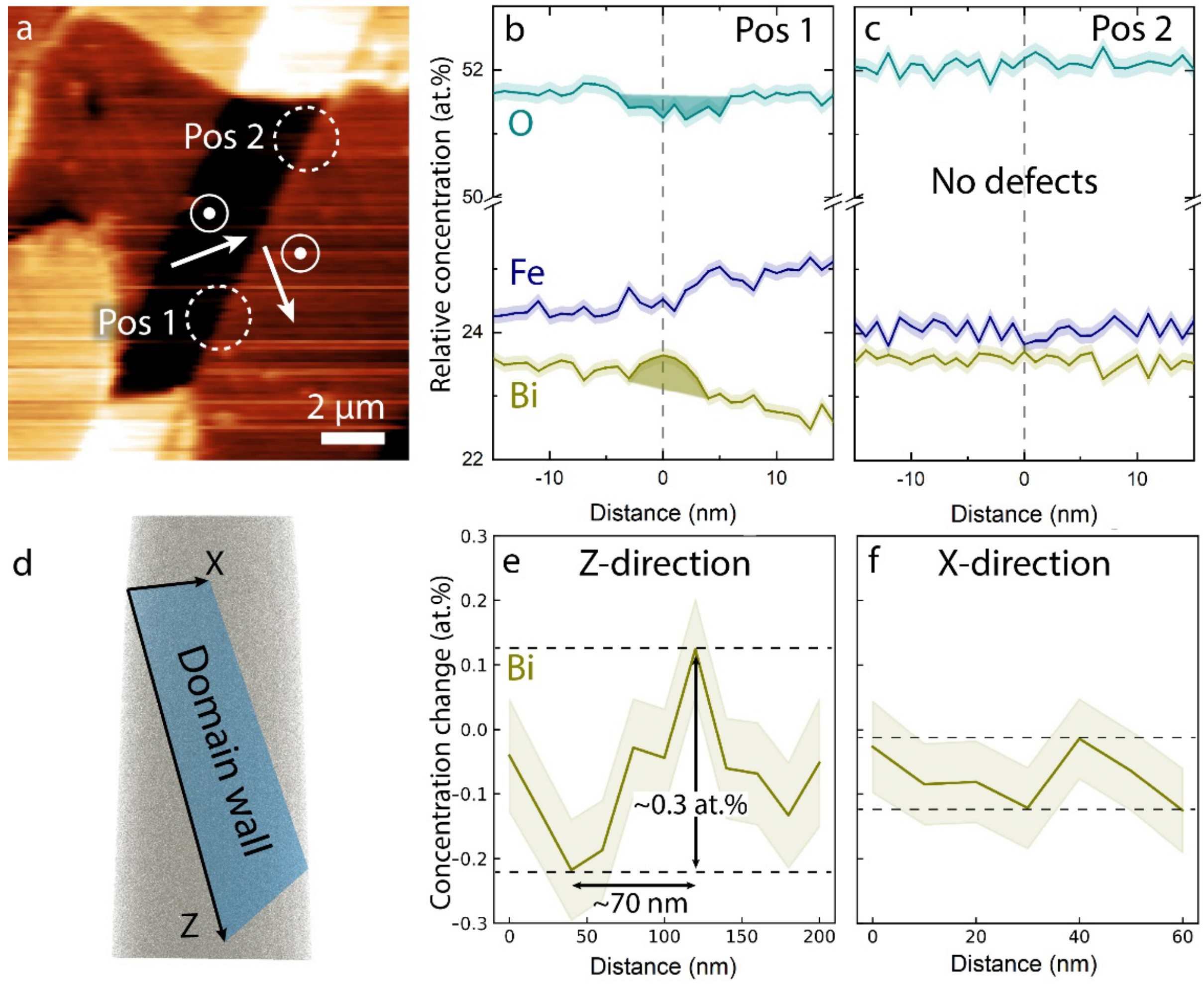

***Fig 3: Correlated SPM-APT experiments. a,*** *PFM scan of a selected site, containing a 109° ferroelectric domain wall. Two positions are marked along the wall, as ① and ②, from where APT specimens are extracted.* ***b,*** *APT analysis from ①. The drop in O concentration and increase in Bi concentration at the domain wall position can be explained as an accumulation of $V_{Fe}$ and $V_O$. Depletion regions are visualized by the shaded regions.* ***c,*** *APT analysis from ②****.*** *No change in chemical composition relative to the domains is resolved at the domain wall position. Correlative TEM data is presented in Supplementary Fig. 4.* ***d,*** *Illustration of the profile geometry in* ***e*** *and* ***f****, where the domain wall plane is visualized in the APT data (Bi distribution). The blue shaded area represents the domain wall plane within an APT reconstruction, with x- and z-direction drawn in.* ***e*** *and* ***f****, The change in Bi concentration at the domain wall is shown within the domain wall plane along the z- and x-direction, respectively, with the maximum observed change indicated in* ***e****. Fe and O changes are presented in Supplementary Fig. 6.*

It is expected that the chemical variations observed within the domain walls locally alter the band structure and lead to intra-bandgap defects states[1]. One example is oxygen vacancies that, in principle, can generate mobile electrons, which co-determine the electron doping level by increasing the n-type conductivity (note that electronic defects are not resolved with APT). This effect is further examined using lattice-resolved phase-field simulations (see Fig. 4a), informed by first-principles density functional theory (DFT) inputs of short-range vacancy pair interactions, as well as long-range vacancy repulsion interactions and the reduction of vacancy formation energy at the domain walls[32]. Starting with a randomly distributed average oxygen vacancy concentration, the simulations demonstrate two main findings: (i) oxygen vacancies tend to accumulate at the 109° domain wall and (ii) these oxygen vacancies spontaneously undergo macroscopic phase separation within the domain wall, creating defect-rich clusters whose sizes depend on the initial average vacancy concentration (see Supplementary Fig. 7)[33,34]. The result indicates that the large-scale inhomogeneities resolved by APT naturally emerge from the underlying defect energetics within the domain wall due to competing short-range attraction and long-range repulsion.

This direct relation between point defect density and electronic conduction provides the opportunity to independently verify emergent fluctuations in defect density along the walls via local transport measurements. On a closer inspection of the current map on the 109° domain walls in Fig. 3a, we indeed observe pronounced spatial variations in conductance as presented in Fig. 4b. The locally measured currents vary between 5 pA and 10 pA on the 100 nm length scale, consistent with the APT results. The respective root-mean-square deviation (RMS) is about 47% of the average conductance value, which is a much larger variation than in the surrounding domains (RMS ≈ 17%). Note that similar fluctuations in conductance have repeatedly been observed for domain walls in the $BiFeO_3$ system[3,15]. Based on our results, we conclude that such fluctuations are a characteristic feature of the $BiFeO_3$ domain walls, originating from their chemical diversity rather than experiment-related effects (e.g., instabilities in the tip-sample contact). The locally enhanced and spatially varying conductance can thus be understood in terms of thermally activated ionization of the segregated defects clusters.

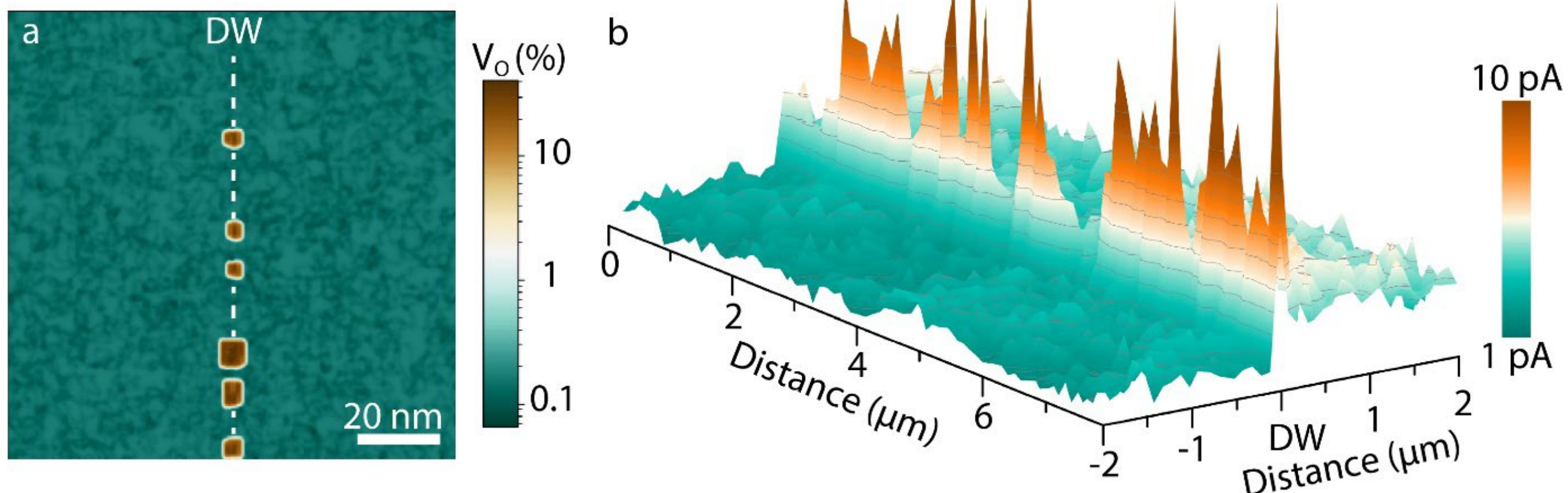

***Fig 4: Spatially varying electronic properties. a,*** *Two-dimensional lattice-resolved phase-field simulation, informed by first-principles energetics, showing macroscopically an inhomogeneous distribution of oxygen vacancies across a charge-neutral domain wall. The distribution is shown by the ratio of missing oxygen atoms to stoichiometric values. The white dashed line mark the domain wall position.* ***b,*** *cAFM measurement of the region shown in Fig. 3a. At the position of the domain wall (DW), the current rises sharply, but varies substantially from position to position along the domain wall.*

## Conclusion

Our results give direct insights into the correlation between the accumulation of point defects and the emergence of enhanced conductivity at ferroelectric domain walls. The 109° domain walls in $BiFeO_3$ act as sinks for point defects with a much greater diversity than previously reported, both in terms of the type and concentration of defects. Between different 109° domain walls, as well as along individual walls, the defect concentration can fluctuate substantially, varying between defect-free and defect-rich sections. This variation in defect concentration is furthermore directly correlated with variations in the local domain wall conduction, which is essential for their electronic response and envisioned applications in domain wall nanoelectronics.

## References

1. Nataf, G. F. *et al.* Domain-wall engineering and topological defects in ferroelectric and ferroelastic materials. *Nat. Rev. Phys.* **2**, 634–648 (2020).

2. Meier, D. *et al.* Anisotropic conductance at improper ferroelectric domain walls. *Nat. Mater.* **11**, 284–288 (2012).

3. Seidel, J. *et al.* Conduction at domain walls in oxide multiferroics. *Nat. Mater.* **8**, 229–234 (2009).

4. Coll, M. *et al.* Towards Oxide Electronics: a Roadmap. *Appl. Surf. Sci.* **482**, 1–93 (2019).

5. Aird, A. & Salje, E. K. H. Sheet superconductivity in twin walls: experimental evidence of $WO_{3-x}$. *J. Phys. Condens. Matter* **10**, L377 (1998).

6. Catalan, G., Seidel, J., Ramesh, R. & Scott, J. F. Domain wall nanoelectronics. *Rev. Mod. Phys.* **84**, 119–156 (2012).

7. Meier, D. & Selbach, S. M. Ferroelectric domain walls for nanotechnology. *Nat. Rev. Mater.* **7**, 157–173 (2022).

8. Gunkel, F., Christensen, D. V., Chen, Y. Z. & Pryds, N. Oxygen vacancies: The (in)visible friend of oxide electronics. *Appl. Phys. Lett.* **116**, 120505 (2020).

9. Guyonnet, J., Gaponenko, I., Gariglio, S. & Paruch, P. Conduction at domain walls in insulating $Pb(Zr_{0.2}Ti_{0.8})O_3$ thin films. *Adv. Mater.* **23**, 5377–5382 (2011).

10. Schröder, M. *et al.* Conducting domain walls in lithium niobate single crystals. *Adv. Funct. Mater.* **22**, 3936–3944 (2012).

11. Catalan, G. & Scott, J. F. Physics and applications of bismuth ferrite. *Adv. Mater.* **21**, 2463–2485 (2009).

12. Zhang, Y. *et al.* Intrinsic Conductance of Domain Walls in $BiFeO_3$. *Adv. Mater.* **31**, 1902099 (2019).

13. Stolichnov, I. *et al.* Persistent conductive footprints of 109° domain walls in bismuth ferrite films. *Appl. Phys. Lett.* **104**, 132902 (2014).

14. Campanini, M. *et al.* Imaging and quantification of charged domain walls in $BiFeO_3$. *Nanoscale* **12**, 9186–9193 (2020).

15. Rojac, T. *et al.* Domain-wall conduction in ferroelectric $BiFeO_3$ controlled by accumulation of charged defects. *Nat. Mater.* **16**, 322–327 (2017).

16. Bencan, A. *et al.* Domain-wall pinning and defect ordering in $BiFeO_3$ probed on the atomic and nanoscale. *Nat. Commun.* **11**, 1–9 (2020).

17. Haselmann, U. *et al.* Negatively Charged In-Plane

and Out-Of-Plane Domain Walls with Oxygen-Vacancy Agglomerations in a Ca-Doped Bismuth-Ferrite Thin Film. *ACS Appl. Electron. Mater.* **3**, 4498–4508 (2021)

18. Liu, Z. *et al.* In-plane charged domain walls with memristive behaviour in a ferroelectric film. *Nature* **613**, 656–661 (2023).

19. Condurache, O., Dražić, G., Sakamoto, N., Rojac, T. & Benčan, A. Atomically resolved structure of step-like uncharged and charged domain walls in polycrystalline $BiFeO_3$. *J. Appl. Phys.* **129**, 054102 (2021).

20. Baek, S. H., Choi, S., Kim, T. L. & Jang, H. W. Domain engineering in $BiFeO_3$ thin films. *Curr. Appl. Phys.* **17**, 688–703 (2017).

21. Seidel, J. *et al.* Domain wall conductivity in La-doped $BiFeO_3$. *Phys. Rev. Lett.* **105**, 197603 (2010).

22. Hunnestad, K. A., Roede, E. D., Van Helvoort, A. T. J. & Meier, D. Characterization of ferroelectric domain walls by scanning electron microscopy. *J. Appl. Phys.* **128**, 191102 (2020).

23. Gault, B., Moody, M., Cairney, J. & Ringer, S. *Atom Probe Tomography*. (Springer, New York, NY, 2012).

24. Gault, B. *et al.* Atom probe tomography. *Nat. Rev. Methods Prim.* **1**, 1–30 (2021).

25. Choi, T. *et al.* Insulating interlocked ferroelectric and structural antiphase domain walls in multiferroic $YMnO_3$. *Nat. Mater.* **9**, 253–258 (2010).

26. Ludacka, U. *et al.* Imaging and structure analysis of ferroelectric domains, domain walls, and vortices by scanning electron diffraction. *npj Comput. Mater.* **10**, 106 (2024).

27. Hunnestad, K. A. *et al.* 3D oxygen vacancy distribution and defect-property relations in an oxide heterostructure. *Nat. Commun.* **15**, 5400 (2024).

28. Hunnestad, K. A. *et al.* Atomic-scale 3D imaging of individual dopant atoms in an oxide semiconductor. *Nat. Commun.* **13**, 4783 (2022).

29. Zhang, Z., Wu, P., Chen, L. & Wang, J. Density functional theory plus U study of vacancy formations in bismuth ferrite. *Appl. Phys. Lett.* **96**, 232906 (2010).

30. Geneste, G., Paillard, C. & Dkhil, B. Polarons, vacancies, vacancy associations, and defect states in multiferroic $BiFeO_3$. *Phys. Rev. B* **99**, (2019).

31. Li, L. *et al.* Control of Domain Structures in Multiferroic Thin Films through Defect Engineering. *Adv. Mater.* **30**, 1802737 (2018).

32. Zhao, G.-D., Dabo, I. & Chen, L.-Q. Thermodynamic stabilization and electronic effects of oxygen vacancies at $BiFeO_3$ neutral ferroelectric domain walls. *Mater. Today Phys.* **58**, 101859 (2025).

33. Shilo, D., Ravichandran, G. & Bhattacharya, K. Investigation of twin-wall structure at the nanometre scale using atomic force microscopy. *Nat. Mater.* **3**, 453–457 (2004).

34. Lee, W. T., Salje, E. K. H. & Bismayer, U. Influence of point defects on the distribution of twin wall widths. *Phys. Rev. B - Condens. Matter Mater. Phys.* **72**, 104116 (2005).

35. Rojac, T. *et al.* Piezoelectric response of $BiFeO_3$ ceramics at elevated temperatures. *Appl. Phys. Lett.* **109**, 042904 (2016).

36. Thompson, K. *et al.* In situ site-specific specimen preparation for atom probe tomography. *Ultramicroscopy* **107**, 131–139 (2007).

37. Koelling, S. *et al.* In-situ observation of non-hemispherical tip shape formation during laser-assisted atom probe tomography. *Journal of Applied Physics* **109**, 104909 (2011).

38. Hunnestad, K. A. *et al.* Correlating laser energy with compositional and atomic-level information of oxides in atom probe tomography. *Mater. Charact.* **203**, 113085 (2023).

39. Farokhipoor, S. & Noheda, B. Conduction through 71°domain walls in $BiFeO_3$ thin films. *Phys. Rev. Lett.* **107**, 127601 (2011).

40. Farokhipoor, S. & Noheda, B. Local conductivity and the role of vacancies around twin walls of (001)-$BiFeO_3$ thin films. *J. Appl. Phys.* **112**, 052003 (2012).

41. Kresse, G. & Furthmüller, J. Efficient iterative schemes for ab initio total-energy calculations using a plane-wave basis set. *Phys. Rev. B - Condens. Matter Mater. Phys.* **54**, 11169–11186 (1996).

42. Kresse, G. & Furthmüller, J. Efficiency of ab-initio total energy calculations for metals and semiconductors using a plane-wave basis set. *Comput. Mater. Sci.* **6**, 15–50 (1996).

43. Huyan, H., Li, L., Addiego, C., Gao, W. & Pan, X. Structures and electronic properties of domain walls in $BiFeO_3$ thin films. *Natl. Sci. Rev.* **6**, 669–683 (2019).

## Methods

### Sample preparation

$BiFeO_3$ ceramics were prepared following the powder-preparation procedure described in Ref. [35]. The starting powder mixture was prepared from $Bi_2O_3$ (99.999%, Alfa Aesar, Ward Mill, MA, USA) and $Fe_2O_3$ (99.998%, Alfa Aesar, Ward Mill, MA, USA). The raw oxides were first pre-milled individually (200 rpm, 4 h in absolute ethanol) and then homogenized (200 rpm, 4 h in absolute ethanol) in a planetary mill (Retsch, PM400, Haan, Germany) in a Bi to Fe molar ratio of 1:1. All milling steps were made in polyethylene vials with yttria-stabilized-zirconia (YSZ) milling balls of 3 mm in diameters. At this stage, to reduce the electrical conductivity, a small amount (0.26 wt. %) of Co was added to the mixture in the form of $Co_3O_4$ (99%, Alfa Aesar), similarly as reported in Ref. [35]. Uniaxially pressed (150 MPa) powder mixtures were reactively sintered at 780 °C for 4 h with 10 K/min of heating and cooling rates. Like reported in a previous study[15], before the analysis, the sample was additionally thermally relaxed with an annealing excursion above Tc, i.e., at 840 °C, with zero hold time, using heating and cooling rate of 5 °C/min.

The $BiFeO_3$ (97 nm) thin film was grown on a $La_{0.7}Sr_{0.3}MnO_3$ (13 nm)/$SrRuO_3$ (1 nm)-buffered $SrTiO_3$ (001) substrate *(CrysTec GmbH)* by pulsed laser deposition using a 248 nm KrF excimer laser. An $SrRuO_3$ layer (1 nm) was first deposited at 700 °C in an oxygen pressure of $1.6\times10^{-2}$ mbar with a repetition rate of 2 Hz and a laser fluence of 0.7 J/cm$^2$. Subsequently, a $La_{0.7}Sr_{0.3}MnO_3$ metallic buffer layer was grown at 700 °C in $1.5\times10^{-2}$ mbar of oxygen with a repetition rate of 2 Hz, 0.7 J/cm$^2$. The $BiFeO_3$ thin film was deposited at 670°C in $1.2\times10^{-1}$ mbar of oxygen using a repetition rate of 8 Hz and a laser fluence of 0.9J/cm$^2$. After the deposition, the sample was cooled to room temperature under an oxygen pressure of $1.2\times10^{-1}$ mbar at a rate of 10 K min$^{-1}$.

**SPM**

AFM, PFM and cAFM measurements were performed using an NTEGRA Prima scanning probe microscope, NT-MDT Spectrum Instruments. cAFM and PFM data were obtained using ASYELEC.01-R2 Ti/Ir coated Si probes (Oxford Instruments, USA), with a measured spring constant and resonance frequency of 1.58 N m-1 and 66.7 kHz, respectively. PFM data was taken at 10 V peak-to-peak, while cAFM data was taken at a constant voltage of 10 V applied to the back electrode.

**SEM-FIB**

A Thermo Fisher Scientific G4 DualBeam UX Focused Ion Beam (FIB) was used to prepare APT specimens from bulk samples, following the procedures described by Ref. [36]. All samples were prepared by a $Ga^+$ ion beam operating at 30 kV. A protection layer and marker of Pt was first deposited to protect the region of interest using electron-beam assisted deposition. Final polishing to remove the protection layer was performed at 2 kV with the $Ga^+$ ion beam. SEM imaging was performed with the same instrument, operated at 3 kV acceleration voltage, and 0.2 nA beam current.

**TEM**

A JEOL 2100F Field Emission Gun (FEG) microscope, operating at 200 kV, was used to perform the needle inspection, DF imaging and orientational analysis of the finished APT specimens. A double tilt holder to orient the needle and a xyz 2k CCD camera was used for imaging and diffraction recording. Specific reflections used for imaging are written in the figure captions. Digital Micrograph, version 3.31, was used for data analysis of the TEM data. The polarization vectors were determined as explained in Supplementary Fig. 3.

Scanning transmission electron microscopy (STEM) was performed using JEOL ARM 200CF operated at 200 kV, with 68-180 mrad collection semi angles. Sample for STEM analysis was prepared by standard methods. The samples were ground and polished to a thickness of 100 µm. After dimpling, they were ultimately thinned to achieve electron transparency using a Gatan PIPS ion-milling system. Data elaboration of the HAADF images is identical to the process reported in Ref. [15].

**APT**

For the APT measurements, a Cameca LEAP 5000XS was used, operated in laser mode. Measurements were performed at cryogenic temperature (25 K). Femtosecond laser pulses triggered the field evaporation, temporarily heating up the APT specimen. Laser pulses with a pulsing frequency of 333 kHz and a pulse energy of 30 pJ were used. The detection rate was set to 0.5 %, meaning that, on average, 5 atoms were detected every 1000 pulse. Raw APT data were reconstructed using AP Suite 6.3, using SEM images to define the radial evolution. See Supplementary Fig. 9 for a representative mass spectrum of $BiFeO_3$. Concentration profiles were created along by counting atoms in a volume, and presenting the relative concentration of each element along one specific direction (integrating the other axes). Note that some profiles are asymmetric, which is common effect of asymmetric heating across interfaces by the laser pulsing.

The defect concentration profiles within the domain wall plane, presented in Fig. 3d-f, were created by taking the chemical composition inside the domain wall volume (assumed to be 10 nm wide), and calculating the difference to the average composition 45 nm away from the domain wall region. Regions directly above and below the domain wall were used to calculate the average value for comparison, to eliminate any changes in composition within the APT specimen due to, for instance, asymmetric laser heating[37] and resulting changes in chemical composition[38].

**Simulation of oxygen vacancy segregation near charge-neutral domain walls**

The inhomogeneous distribution of oxygen vacancies ($V_O$) was modelled surrounding charge-neutral 109° domain walls in $BiFeO_3$ using a lattice-resolved phase-field framework informed by first-principles energetics. Under reducing conditions, neutral $V_O$ (deep donors) were taken as the representative mobile defect[13,21,32,39,40]. A two-dimensional 256 × 256 grid with lattice spacing $a$ = 4.00 Å represented a pseudocubic (001) plane containing one DW introduced as a Gaussian potential well (depth -0.12 eV[32] for the 109° domain wall in $BiFeO_3$ from DFT[41,42]). Oxygen sites were decomposed into three symmetry-equivalent simple-cubic sublattices $i = 1, 2, 3$, and the local concentrations $c_i(\boldsymbol{R})$ (defined at each perovskite unit-cell position $\boldsymbol{R}$) served as the conserved order parameters. Short-range vacancy pair interactions were extracted from DFT total energy differences in a $\sqrt{6}\times\sqrt{8}\times\sqrt{12}$ pseudocubic supercell (120 atoms), capturing on-site, nearest, and next-nearest neighboring couplings. For intra-sublattice interactions, representative values include -0.29 eV along attractive directions and +0.62 eV for repulsive neighbors; next-nearest shells were weaker (~-0.05 eV). Inter-sublattice on-site and nearest neighboring couplings were about +0.56 and +0.02 eV, respectively. Long-range elastic-like repulsion was parameterized as an isotropic kernel fitted to DFT-calculated dependence of vacancy formation energy on concentration (effective strength ~0.13 eV)[32]. Vacancy fields evolve via a multi-component Cahn-Hilliard equation integrated in Fourier space with a semi-implicit scheme. The total average vacancy concentration was fixed at 0.009 ($V_O$ per unit cell) with an initial random noise.

## Acknowledgements

H.S. Søreide and C.A. Hatzoglou are thanked for their support to the APT lab facilities. The Research Council of Norway (RCN) is acknowledged for its support to the Norwegian Micro- and Nano-Fabrication Facility, NorFab, project number 295864, the Norwegian Laboratory for Mineral and Materials Characterization, MiMaC, project number 269842/F50, and the Norwegian Center for Transmission Electron Microscopy, NORTEM (197405/F50). K.A.H. and D.M. thank the Department of Materials Science and Engineering at NTNU for direct financial support. D.M. acknowledges funding from the European Research Council (ERC) under the European Union's Horizon 2020 research and innovation program (Grant Agreement No. 863691). D.M. thanks NTNU for support through the Onsager Fellowship Program and NTNU Stjerneprogrammet. Part of this work (sample preparation, STEM analysis) was supported by the Slovenian Research and Innovation Agency (P2-0105, J7-4637). Andreja Bencan Golob and Goran Drazic are acknowledged for STEM imaging. Maja Makarovic is acknowledged for the $BiFeO_3$ ceramic sample preparation. The theoretical simulation and analysis of phase-field and DFT were supported as part of the Computational Materials Sciences Program funded by the U.S. Department of Energy, Office of Science, Basic Energy Sciences, under Award No. DE-SC0020145 (G.-D.Z., M.-H.Z., T.Y. and L.-Q.C.). L.-Q.C. also appreciate the support from the Donald W. Hamer Foundation through a Hamer Professorship at Penn State. M.T. acknowledges the Swiss National Science Foundation under Project No. 200021 231428 and 200021-236413.

## Author Contributions

K.A.H. prepared the APT samples using FIB, and conducted the APT, TEM, PFM, and cAFM experiments and analyses under supervision from A.T.J.v.H. and D.M. Polycrystalline samples was synthesized under the supervision of T.R.; E.G. prepared the thin film samples, supervised by M.T. Simulations were performed by G.-D.Z., M.-H.Z. and T.Y., supervised by L.-Q.C.; D.M. devised and coordinated the project and, together with K.A.H., wrote the manuscript. All the authors discussed the results and contributed to the final version of the manuscript.

**Supplementary information**

# Chemical heterogeneity at conducting ferroelectric domain walls

**Kasper A. Hunnestad[1,2,*], Guo-Dong Zhao[3], Mao-Hua Zhang[3], Tiannan Yang[3], Elzbieta Gradauskaite[4,5], Antonius T. J. van Helvoort[6], Morgan Trassin[4], Long-Qing Chen[3], Tadej Rojac[7], Dennis Meier[1,8,9,*]**

[1]*Department of Materials Science and Engineering, Norwegian University of Science and Technology (NTNU), 7491 Trondheim, Norway*
[2]Present address: *Department of Electronic Systems, Norwegian University of Science and Technology (NTNU), Trondheim, Norway*
[3]*Department of Materials Science and Engineering, The Pennsylvania State University, University Park, PA, 16802, USA*
[4]*Department of Materials, ETH Zurich, 8093 Zurich, Switzerland*
[5]*Unité Mixte de Physique, CNRS, Thales, Université Paris-Sud, Université Paris-Saclay, Palaiseau, France*
[6]*Department of Physics, Norwegian University of Science and Technology (NTNU), 7491 Trondheim, Norway*
[7]*Electronic Ceramics Department, Jožef Stefan Institute, 1000 Ljubljana, Slovenia.*
[8]*Faculty of Physics and Center for Nanointegration Duisburg-Essen (CENIDE), University of Duisburg-Essen, Duisburg, Germany*
[9]*Research Center Future Energy Materials and Systems, Research Alliance Ruhr, 44780 Bochum, Germany*
*hunnestad.kasper@gmail.com; dennis.meier@uni-due.de

***Supplementary Table 1: Overview of observed defects at ferroelectric domain walls.*** *For defects observed at ferroelectric domain walls in polycrystalline $BiFeO_3$, defect segregation was detected in roughly half of the domain walls. In contrast, ferroelectric domain walls in $BiFeO_3$ thin films mainly do not display any measurable defect segregation.*

| **Compound** | $V_O$ | $V_{Fe}$ | $V_{Bi}$ | **No defects** |
|---|---|---|---|---|
| Polycrystalline $BiFeO_3$ | 4 (36%) | 1 (9%) | 4 (36%) | 7 (64%) |
| Thin film $BiFeO_3$ | 0 | 1 (10%) | 0 | 9 |

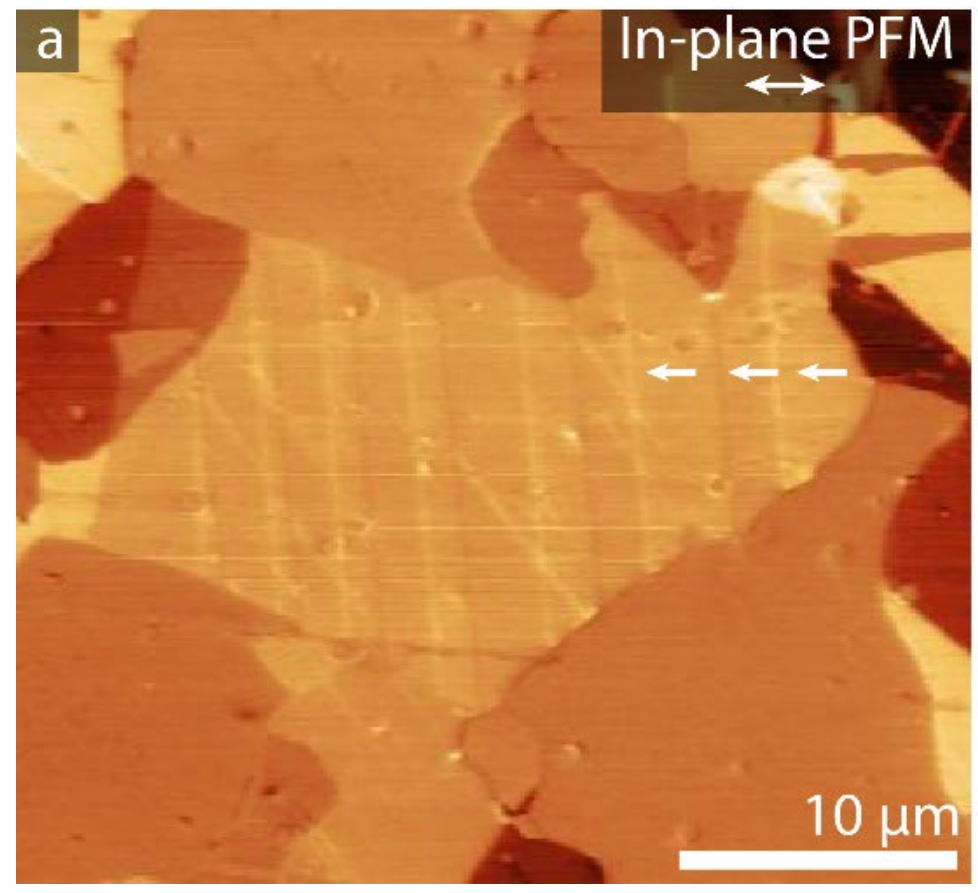

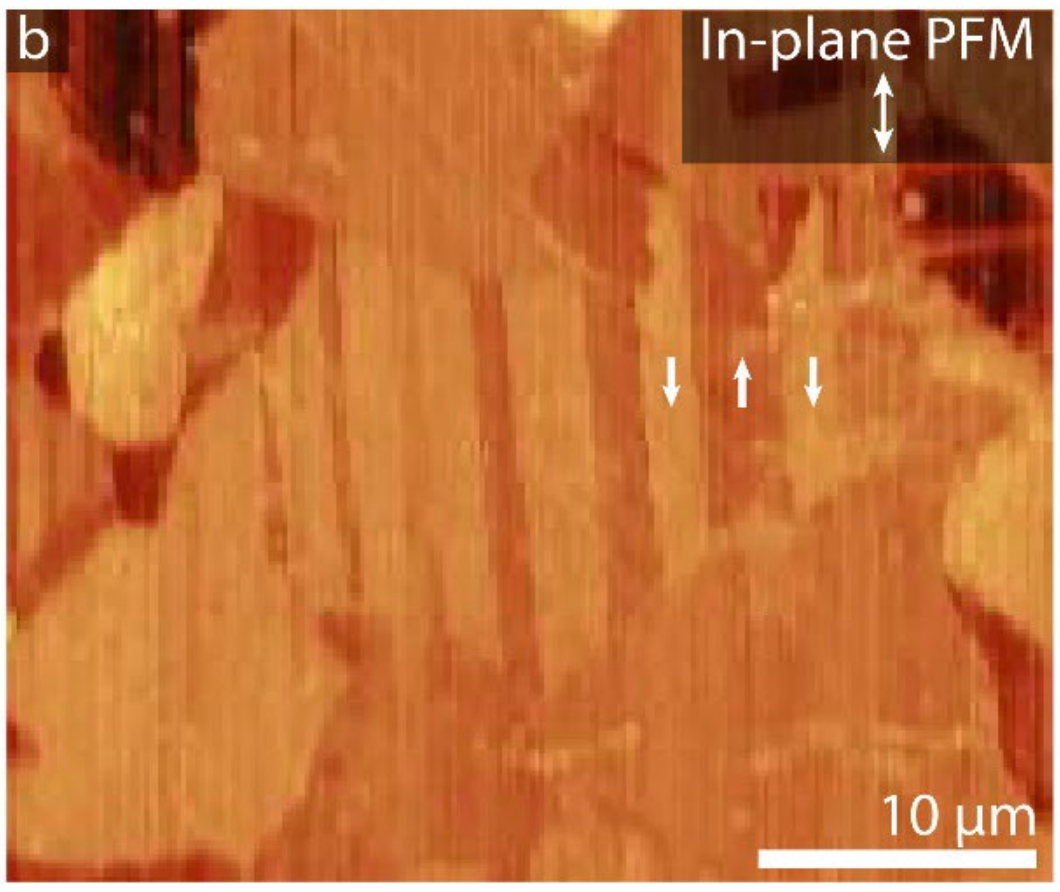

***Supplementary Fig. 1: In-plane PFM images to identify the polarization direction.*** *The in-plane PFM images show the horizontal piezoelectric component, acquired from the region in Fig. 1. The data in **b** are acquired after rotation the sample 90˚ with respect to the orientation in **a**, with the probed axis indicated by the arrow in the top-right corner, and allows for deconvoluting the polarization direction indicated in Fig. 1. In panel **a**, there doesn't appear to be any contrast between the domains within the region of interest, which means that the in-plane polarization direction should be roughly constant and should be predominantly leftwards. From panel **b**, the domains alternate between bright (downwards polarization) and dark (upwards polarization) contrast. Combined with Fig. 1a, the final polarization configuration can thus be determined.*

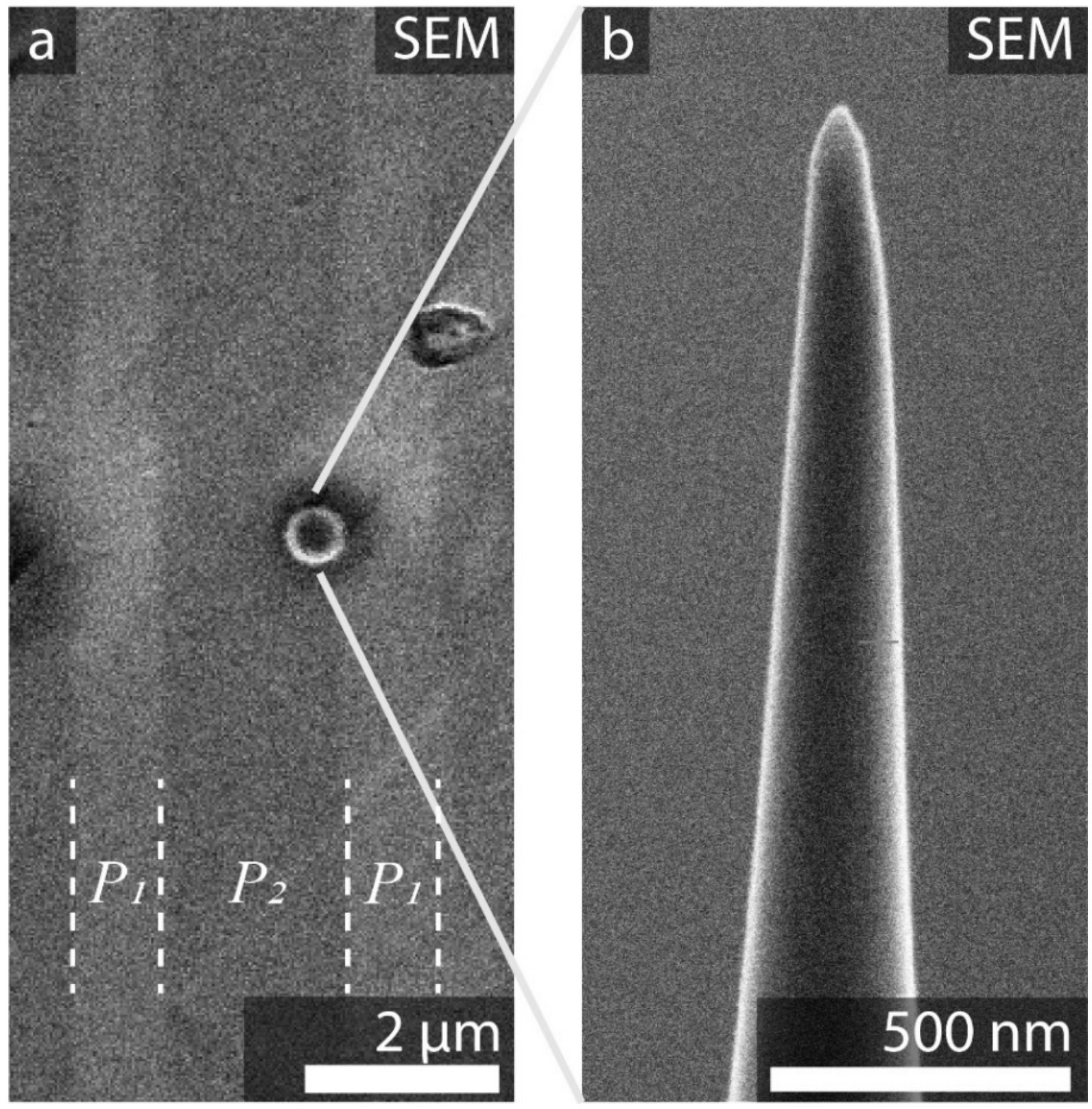

***Supplementary Fig. 2: Domain wall extraction into APT needles. a**, Representative SEM image of an extraction site for APT specimen preparation. Domain contrast is visible in the SEM as alternating bright/dark contrast using secondary electrons[1], and indicated by $P_i$, allowing for precise extraction of APT specimens at the domain wall position. Note that the exact polarization direction is not known at this point. A marker dot made of Pt, deposited using electron beam induced deposition, is visible in the middle of the image and is slightly displaced to the left in the figure to compensate for the sub-surface curvature of the domain wall. The final needle shaped by FIB is shown in **b**.*

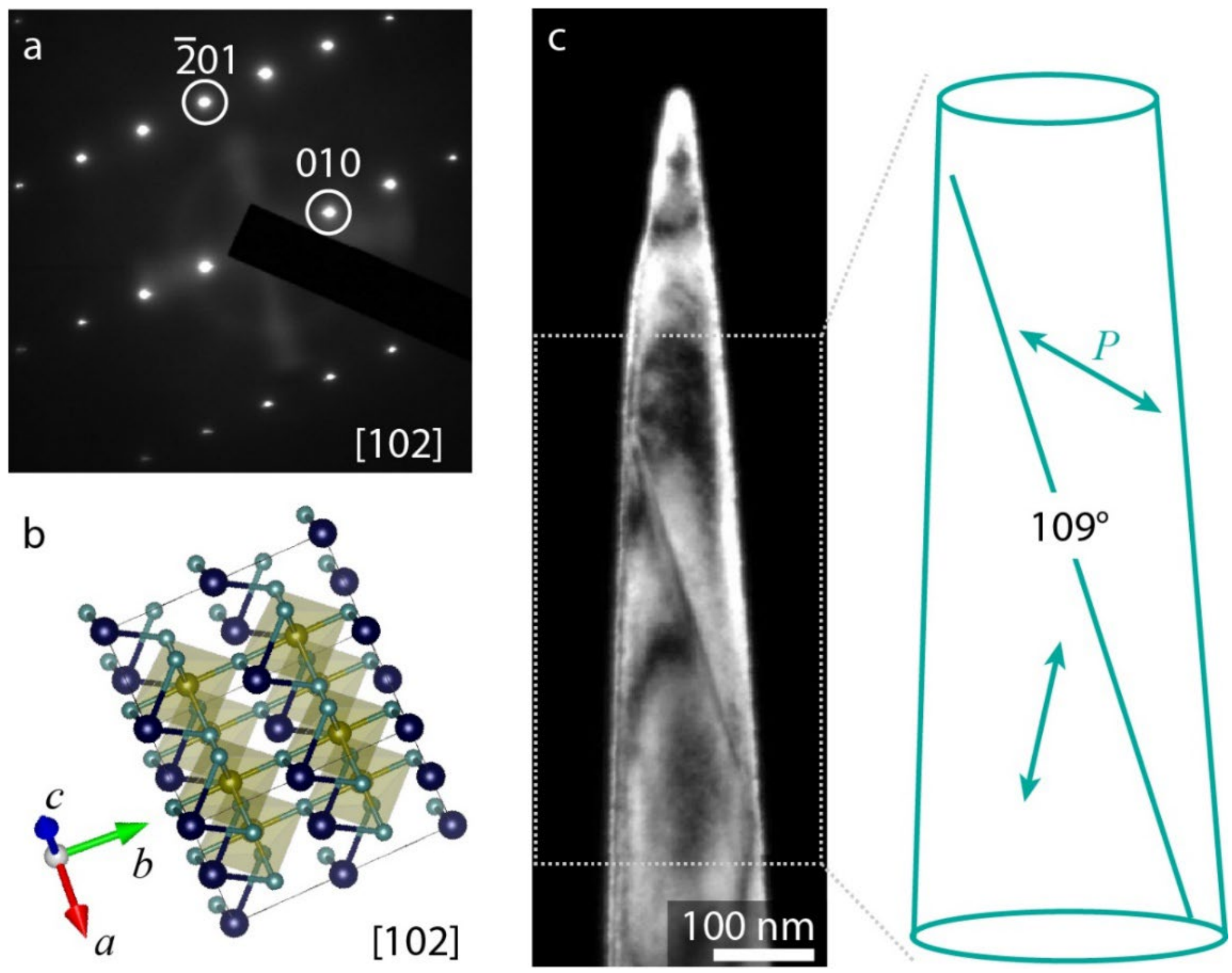

***Supplementary Fig. 3: TEM characterization of the domain wall presented in Fig. 2. a,** SAD pattern of the entire specimen volume. Major reflections are indexed, along with the zone axis. The crystal structure viewing along this orientation is shown in **b,** where the eight possible polar axes extend from the Fe central atoms (yellow) to the corner Bi atoms (blue)**. c,** DF-TEM image of the specimen (using the 0-10 reflection), with a domain wall visible as a straight line abrupting the diffraction contrast. As the domain wall lies in one of the {100} planes and is regular, it can be assumed to be an uncharged 109° domain wallwall[2]. Combined with the information from the PFM data (Fig. 1 and Supplementary Fig. 1), and how the polarization is known to change across a 109° domain wall, the polarization vectors can be determined, as shown schematically on the right.*

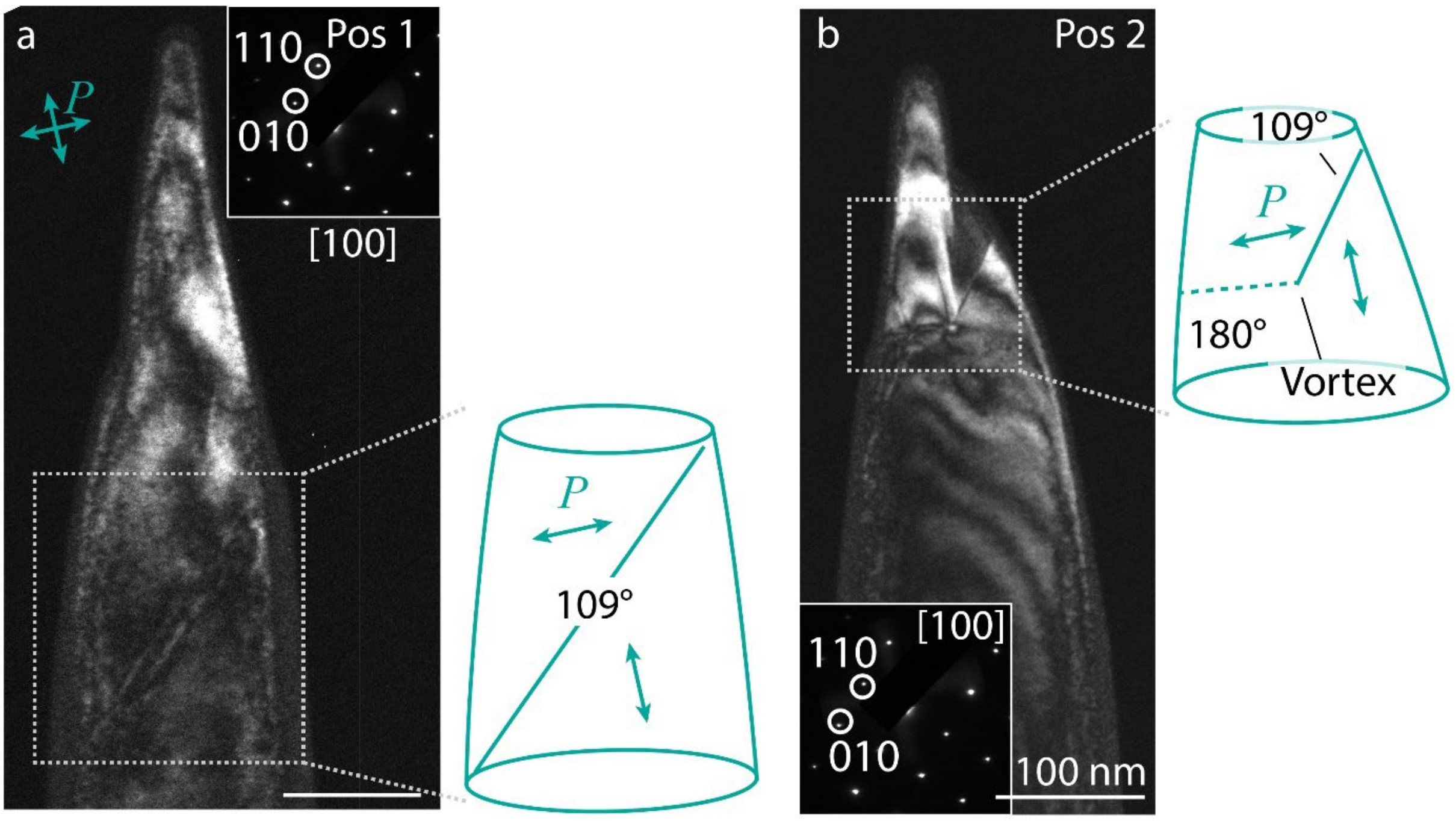

***Supplementary Fig. 4: TEM characterization of the ferroelectric domain walls presented in Fig. 3. a,*** *DF-TEM image of an APT specimen containing a 109° domain wall, extracted from position 1 (Pos 1) in Fig. 3. The crystal orientation is determined from the SAD pattern given in the inset, and the possible polar axes are illustrated by the cyan arrows. A schematic representation of the domain wall and polar configuration is shown next to the DF-TEM image.* ***b,*** *DF-TEM image for position 2 (Pos 2). The specimen contains a domain wall vortex in addition to the 109° domain wall.*

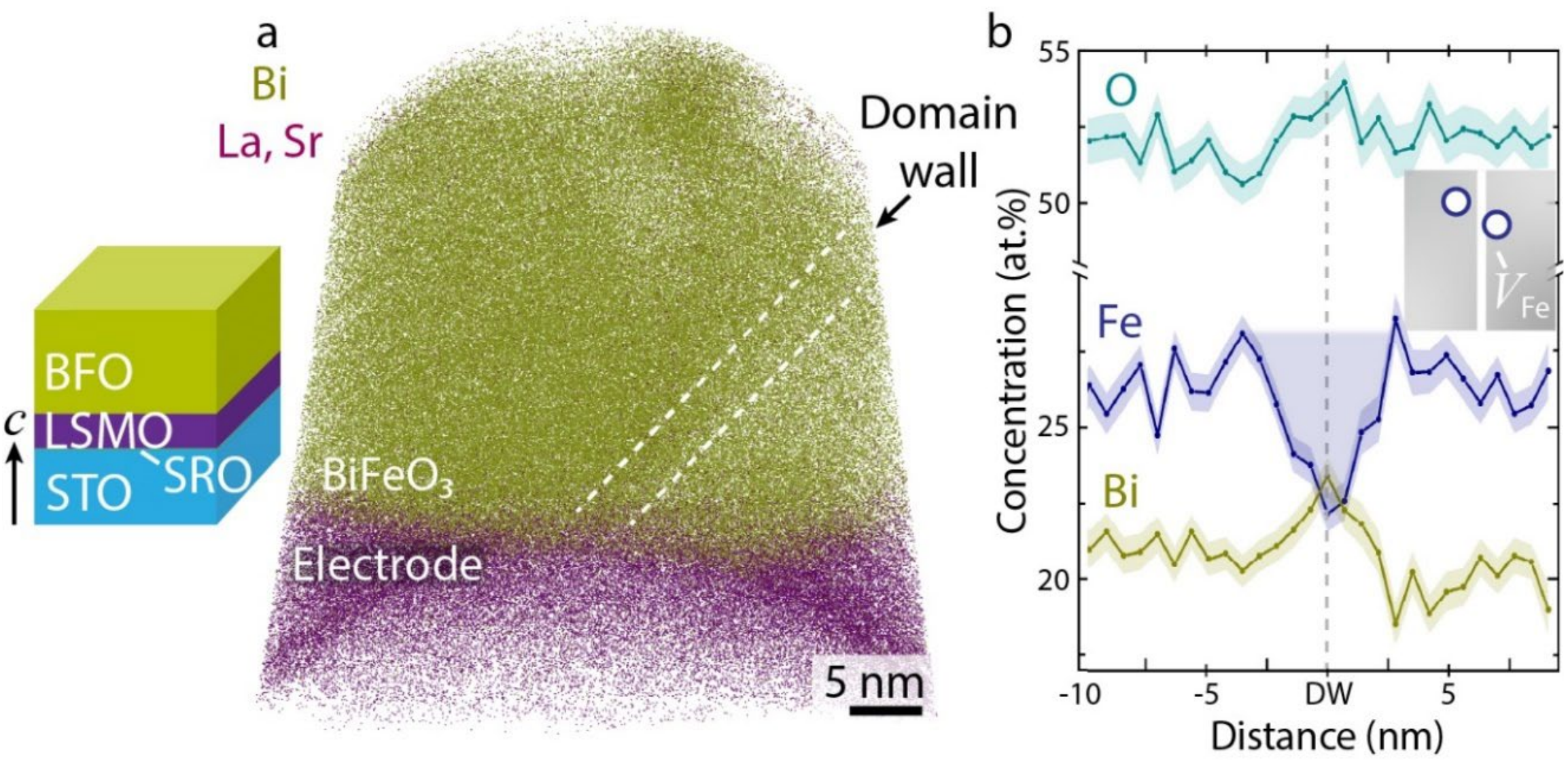

***Supplementary Fig. 5: APT analysis for $BiFeO_3$ thin films. a****, 3D APT reconstruction of the $BiFeO_3$ thin film, grown on an $LaSrMnO_3$ (LSMO) electrode, and an $SrTiO_3$ substrate and parts of the electrode. A schematic representation of the thin film structure is shown next to the reconstruction. The domain wall can be seen to separate two regions with different density, and is indicated by the dashed white lines. The angle of the domain wall is tilted 45° with respect to the* c*-axis, consistent with a 71° domain wall.* ***b****, A profile of the chemical composition across the domain wall, showing a substantial drop in Fe concentration. The change in chemical composition can be explained based on $V_{Fe}$, schematically drawn in the inset. PFM imaging of the sample surface is shown in Supplementary Fig. 8. In contrast to the polycrystals, 71° domain walls are the most prevalent here. More importantly, the domain walls are more densely packed, with an average distance between domain walls of less than 200 nm. For APT sample preparation, this has the advantage that every APT sample can be assumed to contain a domain wall, without the need for confirmation by TEM. Deducing that a domain wall exists within the dataset is thus based on the observation of any planar chemical fluctuation.*

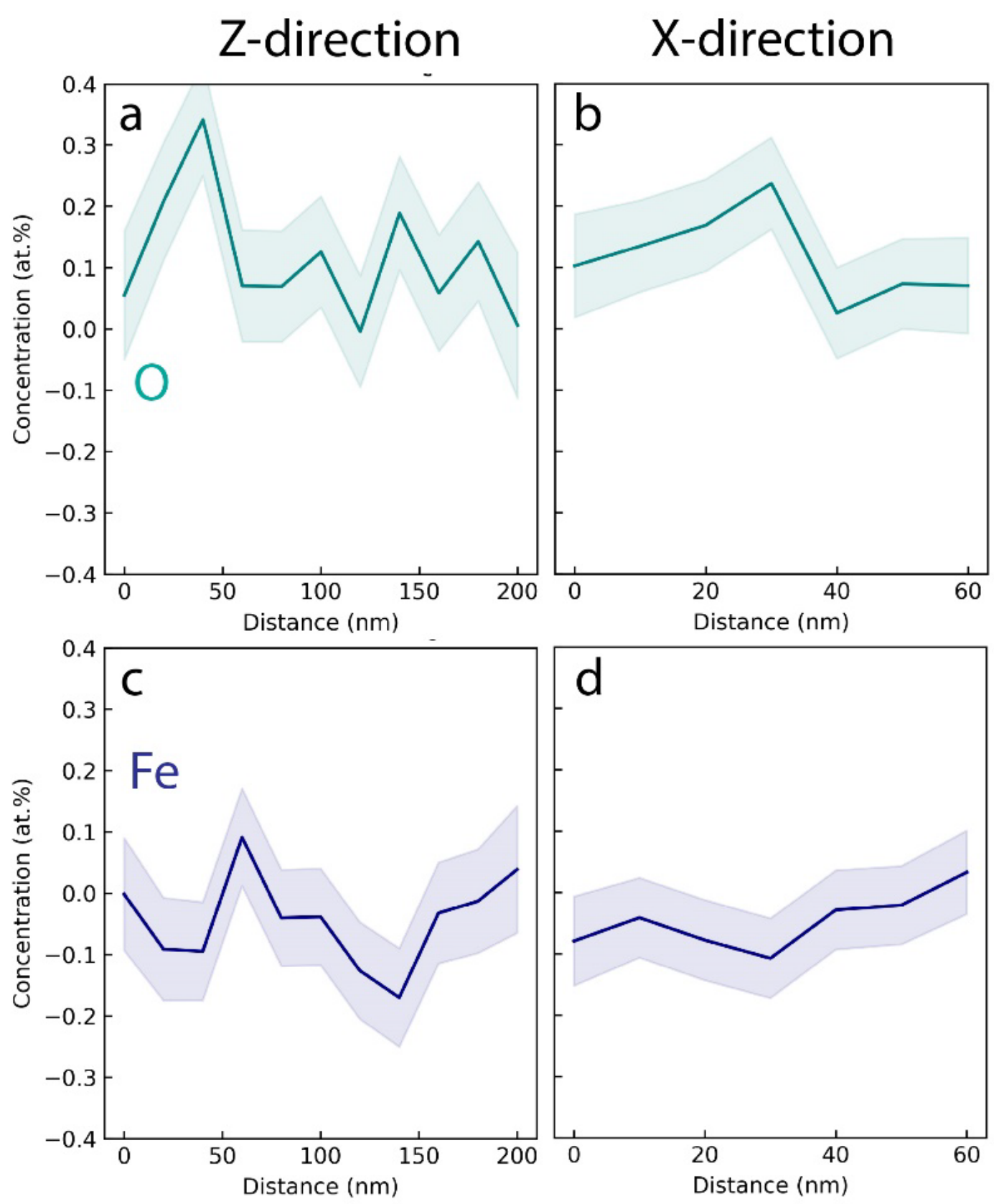

***Supplementary Fig. 6: In-plane composition of a domain wall.*** *Panels* ***a*** *and* ***b*** *show the oxygen concentration changes from bulk within the domain wall plane from the same sample as presented in Fig. 3e and f, where as panels* ***c*** *and* ***d*** *show the Fe concentration changes. All profiles are created in a similar fashion as Fig. 3e and f, described in the Methods.*

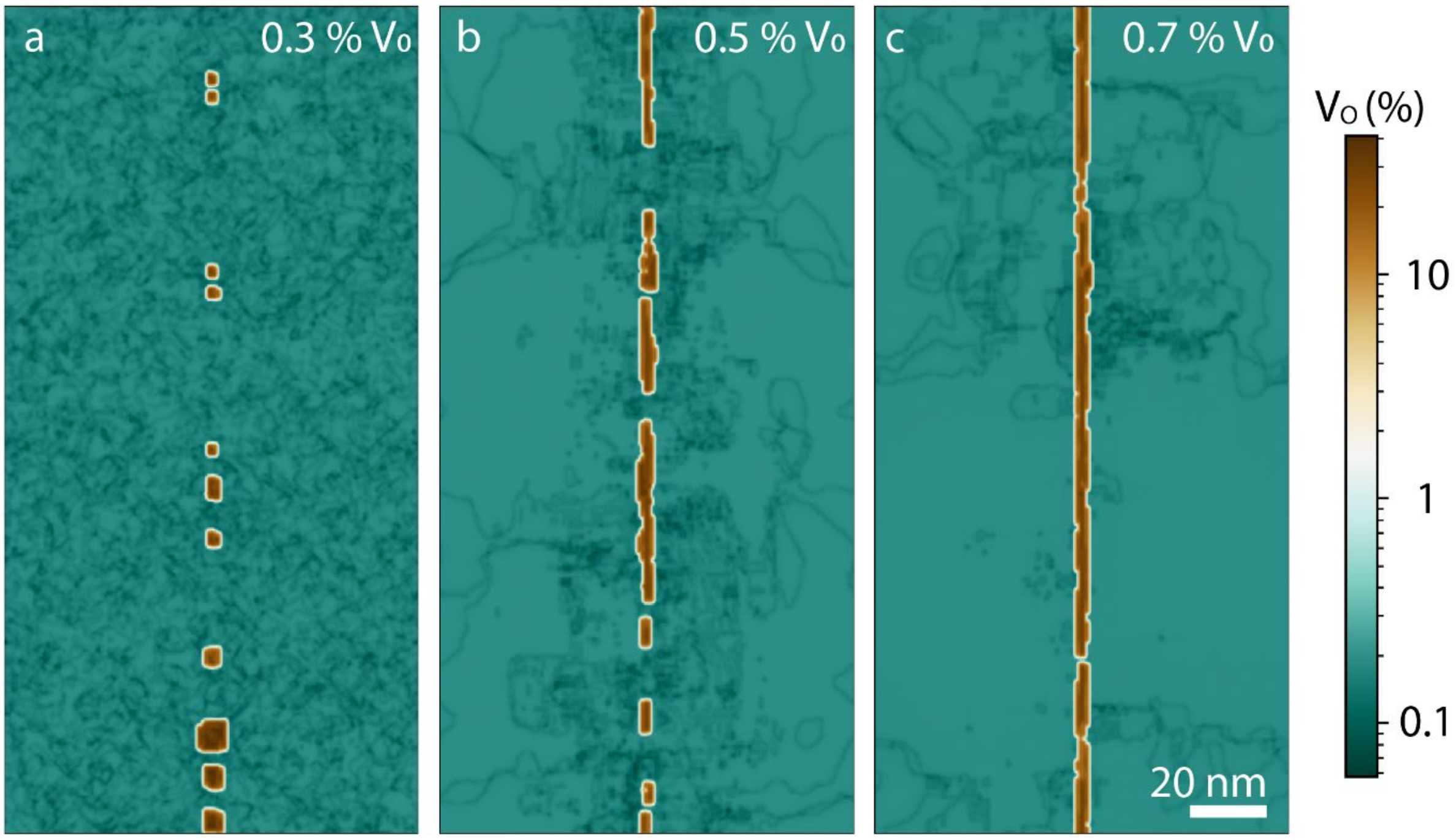

***Supplementary Fig. 7: Extended phase-field simulations of oxygen vacancy distributions from different starting conditions.*** *Two-dimensional lattice-resolved phase-field simulations informed by first-principles energetics, showing macroscopically inhomogeneous distributions of oxygen vacancies across charge-neutral domain walls. The distribution is shown by the percentage of missing oxygen atoms. The domain wall position is in the center of each panel. Panels **a**-**c** represent the oxygen vacancy distribution resulting from different initial bulk concentrations of oxygen vacancies, indicated in the top-right corner.*

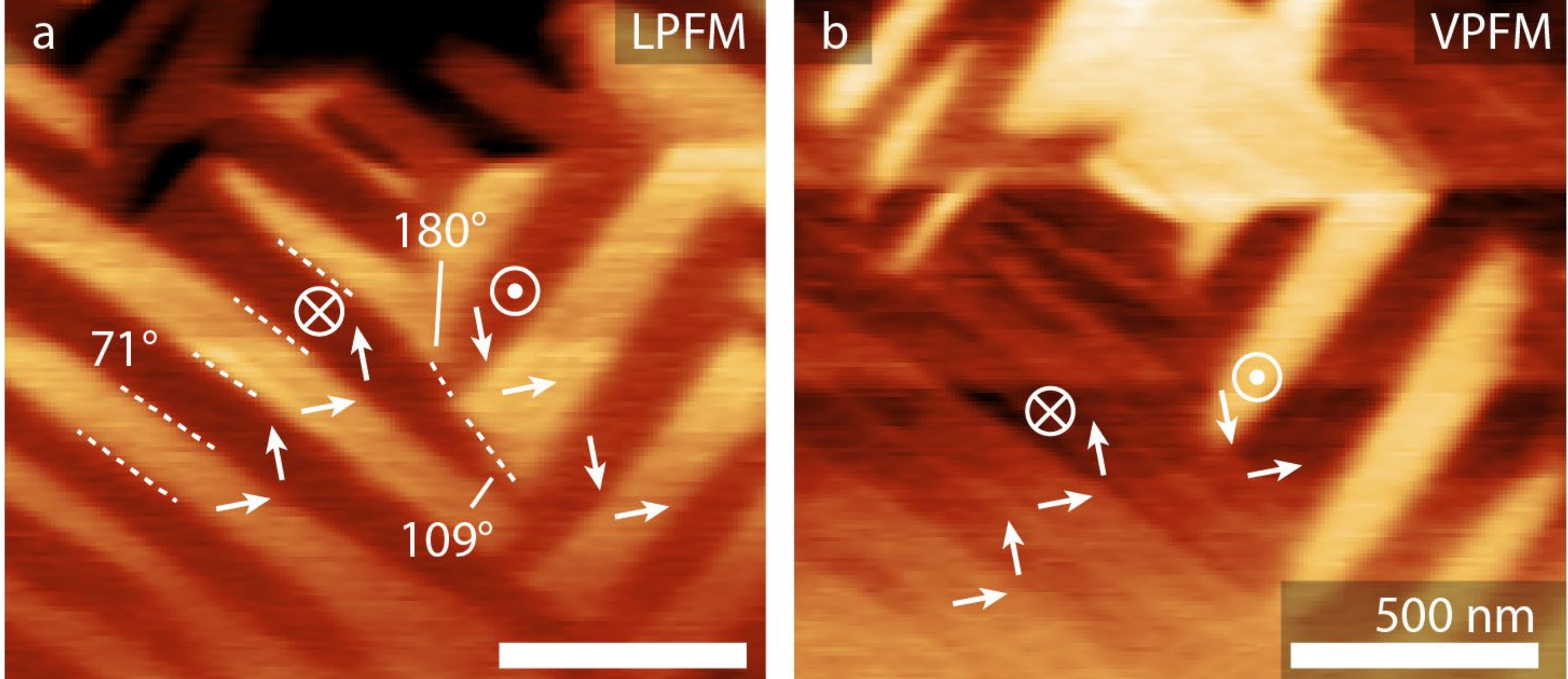

***Supplementary Fig. 8: Domain structure of $BiFeO_3$ thin films. a***, *Lateral PFM (LPFM) and **b**, vertical PFM (VPFM) of the $BiFeO_3$ thin film, displaying a high-density of 71° stripe domain walls.*

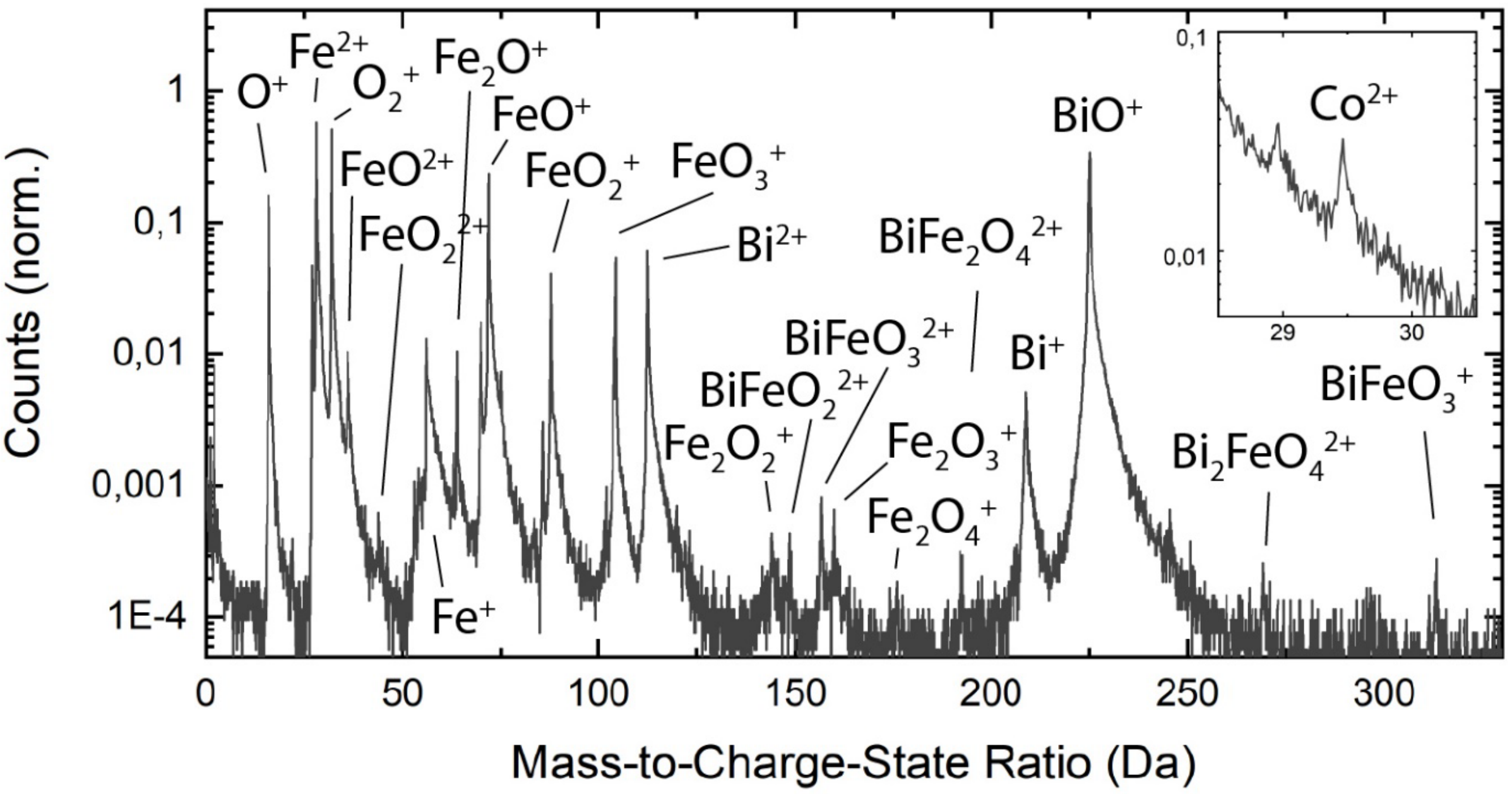

***Supplementary Fig. 9: Mass spectrum of $BiFeO_3$.*** *The ionic species used for the reconstruction are labelled, including the solute Co dopants in the inset. The broader Bi-species peaks are likely to cause a general (microscopically constant) underestimation of the Bi concentration.*

***Supplementary video 1: T****he Bi distribution of Fig. 2b, rotating around the Z-axis. The domain wall can briefly be seen going diagonally across the sample.*

## SI References

1. Hunnestad, K. A., Roede, E. D., Van Helvoort, A. T. J. & Meier, D. Characterization of ferroelectric domain walls by scanning electron microscopy. *J. Appl. Phys.* **128**, 191102 (2020).
2. Huyan, H., Li, L., Addiego, C., Gao, W. & Pan, X. Structures and electronic properties of domain walls in BiFeO3 thin films. *Natl. Sci. Rev.* **6**, 669–683 (2019).